\documentclass[conference,10pt]{IEEEtran}
\IEEEoverridecommandlockouts

\usepackage{stmaryrd}
\usepackage{amsmath}
\allowdisplaybreaks
\usepackage{amssymb}
\usepackage{graphicx}
\usepackage{todonotes}
\usepackage{cite}
\usepackage{algorithmic}
\usepackage{algorithm}
\usepackage{mathtools}
\usepackage{caption}
\usepackage{subcaption}
\usepackage{float}
\usepackage{comment}
\usepackage{makecell}
\usepackage{xcolor}
\usepackage{multirow}
\usepackage{bm}
\usepackage{bbm}
\usepackage{glossaries}

\newcommand{\vect}[1]{\boldsymbol{\mathrm{#1}}}
\newcommand{\mat}[1]{\boldsymbol{\mathrm{#1}}}

\newcommand{\prob}[1][]{
\ifthenelse{\isempty{#1}}%
      {\ensuremath{P}}%
    {\ensuremath{P\left\(#1\right\)}}%
}

\def\defeq{\triangleq}

\newcommand{\transp}{{\sf T}}

\def\bv{{\vect{v}}}

\def\bx{{\vect{x}}}
\def\by{{\vect{y}}}

\def\b0{{\vect{0}}}

\def\bD{{\mat{D}}}

\def\bH{{\mat{H}}}
\def\bI{{\mat{I}}}

\def\bX{{\mat{X}}}

\def\cG{{\mathcal{G}}}

\def\C{{\mathbb{C}}}

\newtheorem{remark}{Remark}

\newenvironment{proof outline}{\paragraph*{Proof Outline}}{\hfill$\IEEEQEDopen$}

\def\Circ{\textup{Circ}}

\newacronym{cfr}{CFR}{channel frequency response}
\newacronym{cir}{CIR}{channel impulse response}
\newacronym{lti}{LTI}{linear time-invariant}
\newacronym{bibo}{BIBO}{bounded-input bounded-output}
\newacronym{mimo}{MIMO}{multiple-input multiple-output}
\newacronym{dft}{DFT}{discrete Fourier transform}
\newacronym{los}{LoS}{line-of-sight}
\newacronym{lhp}{LHP}{left half-plane}
\newacronym{nr}{NR}{New Radio}

\makeglossaries

\begingroup
\renewcommand*{\arraystretch}{1.5}
\endgroup

\makeatletter
\renewcommand*\env@matrix[1][\arraystretch]{%
  \edef\arraystretch{#1}%
  \hskip -\arraycolsep
  \let\@ifnextchar\new@ifnextchar
  \array{*\c@MaxMatrixCols c}}
\makeatother

\begin{document}

\title{Stability Analysis of Interacting Wireless Repeaters
\thanks{
The authors are with the Department of Electrical Engineering (ISY), Link\"oping University, 58183 Link\"oping, Sweden (email: jianan.bai@liu.se, erik.g.larsson@liu.se). 
This work was supported in part by Excellence Center at Link\"oping-Lund in Information Technology (ELLIIT), and by the Knut and Alice Wallenberg (KAW) foundation. 
}
}

\author{Erik G. Larsson and Jianan Bai}

\maketitle

\begin{abstract}

We consider a wireless network with multiple single-antenna repeaters that amplify and instantaneously re-transmit the signals they receive to improve the channel rank and system coverage.
Due to the positive feedback formed by inter-repeater interference, stability could become a critical issue.
We investigate the problem of determining the maximum amplification gain that the repeaters can use without breaking the system stability. 
Specifically, we obtain a  bound by using the Gershgorin disc theorem, which reveals that the maximum amplification gain is restricted by the sum of channel amplitude gains.
We show by case studies the usefulness of the so-obtained bound and provide insights on how the repeaters should be deployed.
	
\end{abstract}


\section{Introduction}

Wireless networks are expected to serve users reliably, regardless of their locations.
However, in practice, network coverage is inevitably restricted by signal attenuation, shadowing, and blockage, which are important effects, especially in higher frequency bands in 5G \gls{nr}.
In addition, in many environments, the low rank of the channel limits the number of streams that can be multiplexed.
To compensate for these effects, one can either choose to deploy more powerful macro base stations equipped with massive-antenna technology, or to densify the network by deploying additional small cells or use cell-free schemes.
Although effective, these schemes are costly and may not offer a cost-effective solution.

Wireless repeaters are devices with instantaneous amplify-and-forward operation in full-duplex. 
Compared with the deployment of additional base stations, repeaters can have extremely attractive form factors, 
they require no backhaul or phase synchronization, and they can operate with very low power consumption. 
These advantages make repeaters easy and inexpensive to deploy.
Repeaters have been widely deployed in 2G, 3G, and 4G networks to establish coverage especially in tunnels, and have been consistently shown to provide capacity and coverage improvements in various scenarios~\cite{wen2024shaping,patwary2005capacity,sharma2015repeater,garcia2007enhanced,tsai2010capacity,ayoubi2023network,leone2022towards,ma2015channel}.
In the context of 5G NR and the upcoming 6G, significant efforts have been made to standardize network-controlled repeaters in 3GPP \cite{TS38-106}, which enable more versatile functionality through the use of control information.

Repeaters should not be confused with relays.
A relay receives a waveform of some duration, and re-transmits it (with or without decoding).
A repeater \emph{instantaneously} retransmits, with a delay of a few 100 ns at most, making it act as an ordinary channel scatterer, but with amplification \cite{willhammar2024achieving}. 
As long as the repeater is reciprocal, it will be transparent to the
network in a TDD multiuser MIMO system \cite{willhammar2024achieving,larsson2024reciprocity}.
An important design aspect of repeaters is to keep self-interference under control. This self-interference
can be a limiting factor for the repeater gain.
For   single-antenna repeaters,  it is  the antenna-interface circuit,
and for dual-antenna repeaters also the inter-antenna distance, that determine the isolation between the transmit and receive paths and hence determine the maximum gain \cite{willhammar2024achieving,chen2021survey}.

To provide better coverage extension, swarms of repeaters have been envisioned to be densely deployed, with large amplification gains. 
However, due to the broadcast nature of wireless channels and the full-duplex operation of repeaters, inter-repeater interference forms a positive feedback loop that could drive the repeater amplifier into saturation and cause it to malfunction. 
Preventing such destructive positive feedback is crucial for the successful deployment of repeater swarms.
Two fundamental questions need to be addressed:
\begin{itemize}
	\item For a particular deployment of repeaters, what is the maximum amplification gain that can be used without causing destructive positive feedback?
	\item How densely should repeaters be deployed to achieve the maximum coverage extension?
\end{itemize}

\emph{Technical contributions:}
We model the interaction between repeaters, and the positive feedback phenomenon, as an input-output stability problem from a system theory viewpoint.
Particularly, when all repeaters employ the same amplification gain and this gain is continuously increased from zero, we mathematically characterize the maximum amplification gain as the transition point at which the system  becomes unstable.
Given the difficulty in finding the exact value of this maximum amplification gain for general scenarios, we provide a bound on it by using the Gershgorin disc theorem.
Through case studies and numerical results, we show that the obtained lower bound accurately captures the stability transition point.
We also provide insights on how densely the repeaters should be deployed by considering both the stability and power constraints.

\emph{Notations:}
We use $\hat{x}(s)$ as the Laplace transform of the time-domain representation $x(t)$. 
The frequency-domain representation at the angular frequency $\omega$ is therefore $\hat{x}(j\omega)$.
Vectors are denoted by boldface lowercase letters, $\bx$, and matrices by boldface uppercase letters, $\bX$.
A diagonal matrix with $\bx$ on its diagonal is denoted by $\bD_\bx$.
For a positive integer $N$, we use $[N]$ to represent the set $\{1,\cdots,N\}$.

\section{System Model}

\begin{figure}
	\centering
	\includegraphics[width=5.5cm]{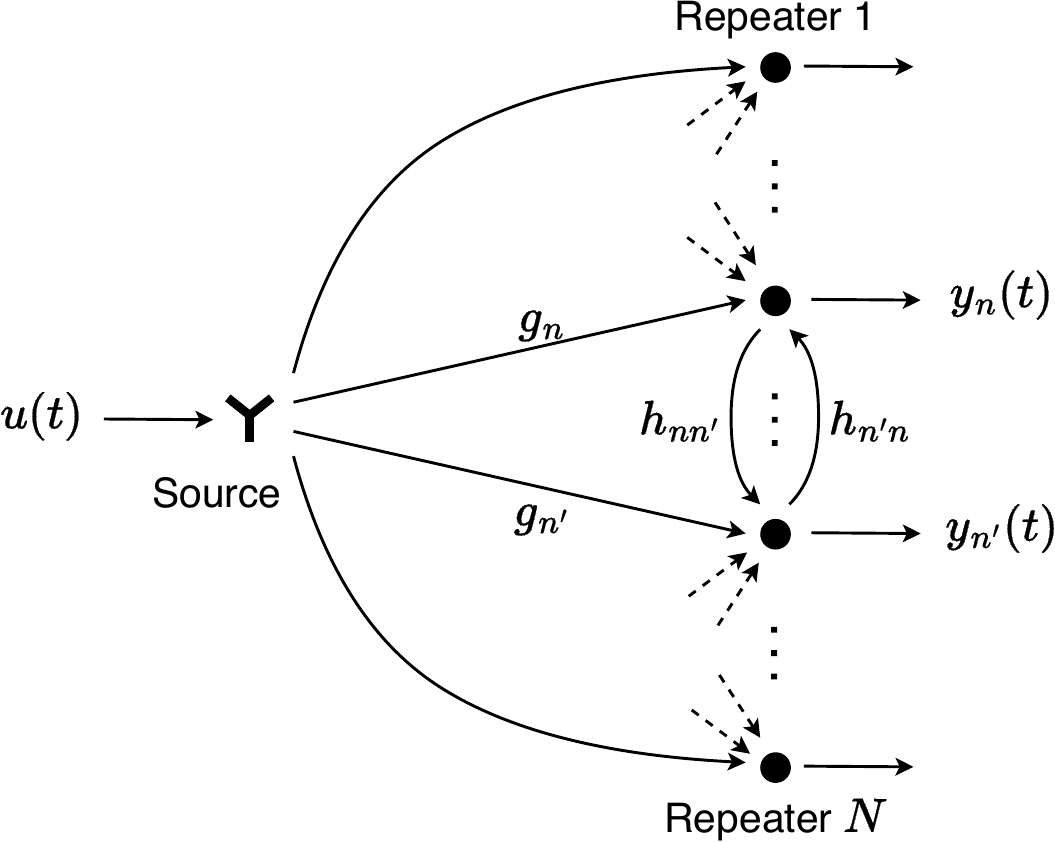}
	\caption{$N$ interacting single-antenna wireless repeaters.}
	\vspace*{-.7cm}
	\label{fig: repeater-model}
\end{figure}

We consider a wireless network with one source and $N$ repeaters, each with a single antenna. 
The repeaters are assumed to be ideal signal amplifiers with the same amplification gain $\alpha > 0$ and without any induced delay.
All wireless channels are assumed to be \gls{lti}.
The \gls{cir} from the source to the $n$-th repeater is denoted by $g_n(t)$.
The \gls{cir} from the $n'$-th repeater to the $n$-th repeater is $h_{nn'}(t)$.
The channels are assumed to be reciprocal, i.e., $h_{nn'}(t) = h_{n'n}(t), \forall n,n'\in[N]$.
Also, we assume that  self-interference is removed by the antenna-interface circuit or by using
sufficient antenna separation, i.e., $h_{nn}(t) = 0$.
See Fig. \ref{fig: repeater-model} for an illustration.

When the source transmits a signal $u(t)$, the output from repeater $n$ is (noise is ignored)
\begin{equation}
\label{eq: signal in the time domain}
	y_n(t) = \alpha \bigg(g_n(t)*u(t) + \sum_{n'\neq n} h_{nn'}(t)*y_{n'}(t) \bigg),
\end{equation}
where ``$*$'' represents convolution.\footnote{Note that all signals (including the signal from the source) are
amplified by the repeater and hence scaled by $\alpha$ in (\ref{eq: signal in the time domain}).}
Notice that the output consists of both the signal emitted from the source and the signals echoed among the repeaters.
To facilitate the analysis, we represent the system in the Laplace domain:
\begin{equation}
	\hat{y}_n(s) = \alpha \bigg(\hat{g}_n(s)\hat{u}(s) + \sum_{n'\neq n} \hat{h}_{nn'}(s) \hat{y}_{n'}(s)\bigg),
\end{equation}
where $\hat{u}(s)$, $\hat{y}(s)$, $\hat{g}(s)$ and $\hat{h}_{nn'}(s)$ are the Laplace transforms of $x_n(t)$, $y_n(t)$, $g_n(t)$, and $h_{nn'}(t)$, respectively. (For brevity, the Laplace variable $s$ will be omitted henceforth.)

\begin{figure}
	\centering
	\includegraphics[width=5.5cm]{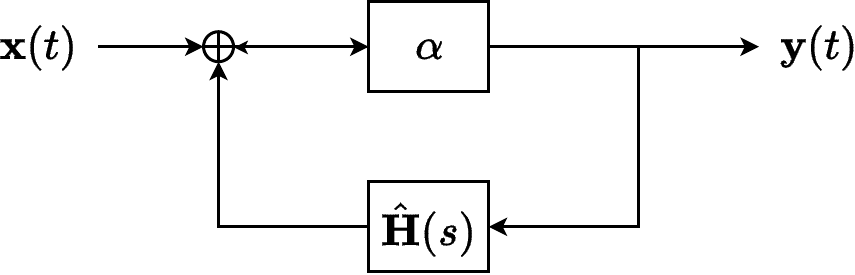}
	\caption{Block diagram corresponding to \eqref{eq: mimo system}.}
	\label{fig: block diagram}
	\vspace{-0.7cm}
\end{figure}

By defining the vectors $\hat{\bx} \defeq  [\hat{x}_1,\cdots,\hat{x}_N]^\transp$ with $\hat{x}_n \defeq \hat{g}_n\hat{u}$, $\hat{\by} \defeq [\hat{y}_1,\cdots,\hat{y}_N]^\transp$, and the channel transfer function matrix
\begin{equation}
	\hat{\bH} \defeq 
	\begin{bmatrix}
		\hat{h}_{11} & \cdots & \hat{h}_{1N}\\
		\vdots & \ddots & \vdots \\
		\hat{h}_{N1} & \cdots & \hat{h}_{NN}
	\end{bmatrix},
\end{equation}
the input-output relationship of all repeaters is given by
${ 
	\hat{\by} = \alpha(\hat{\bx} + \hat{\bH} \hat{\by}),
}$ 
or equivalently
\begin{equation} 
\label{eq: mimo system}
	\hat{\by} = \alpha (\bI - \alpha \hat{\bH})^{-1} \hat{\bx},
\end{equation}
which represents a positive-feedback \gls{mimo}\footnote{The system theory term ``\gls{mimo}'' here should not be confused with MIMO in multi-antenna wireless communications.} system with the block diagram in Fig. \ref{fig: block diagram}.

\section{Stability Analysis}

To improve the coverage of a wireless network, a larger amplification gain $\alpha$ is preferred. 
However, as the interaction of repeaters forms a positive feedback loop, closed-loop stability becomes a critical design aspect. 
In this section, we formally characterize the system stability and provide a lower bound on the maximum amplification gain that can be used by the repeaters without making the system unstable.

\subsection{Maximum Stable Amplification Gain}

When all repeaters are turned off, i.e., $\alpha = 0$, there is no positive feedback, and therefore the system is stable. 
As we continuously increase the amplification gains of the repeaters, the system starts to become unstable for some $\alpha = \alpha_{\max} > 0$. 
This particular value of $\alpha$ is referred to as the \emph{maximum stable} amplification gain in the sense that one should not employ an amplification gain larger than $\alpha_{\max}$ to keep the system stable.
We aim to find $\alpha_{\max}$ for arbitrary deployments of repeaters so that their amplification gains can be properly configured without affecting system stability.
Furthermore, $\alpha_{\max}$ determines the best possible coverage improvement one can expect by deploying repeaters.

When $\alpha$ continuously grows from zero, $\alpha_{\max}$ is the \emph{transition point} at which some pole(s) of the closed-loop transition function matrix $\alpha(\bI - \alpha\hat{\bH})^{-1}$ come across the $j\omega$-axis.
The maximum amplification gain can be formally defined as
\begin{equation}
\label{eq: maximum amplification gain}
	\alpha_{\max} \defeq \min \left\{\alpha>0: \inf_{\omega} |\det(\bI - \alpha\hat{\bH}(j\omega))| = 0 \right\}.
\end{equation}
In \eqref{eq: maximum amplification gain}, we look for the \emph{minimum} 
positive value of $\alpha$ such that $\bI-\alpha\hat{\bH}(j\omega)$ becomes singular for \emph{some} frequency $\omega$.
Note that  the determinant in  \eqref{eq: maximum amplification gain} is zero precisely when there is a   zero eigenvalue. Note also that the eigenvalues are complex since $\hat{\bH}(j\omega)$ is symmetric but not
Hermitean.
Furthermore, note that  the determinant is only zero for   the ``critical values''  of $\alpha$ corresponding to
zero eigenvalues. This means that if $\alpha$ is increased without bound, the determinant will be
nonzero. However, what matters for the stability is the \emph{smallest} value of $\alpha$ that makes the
determinant zero.

To find $\alpha_{\max}$ using \eqref{eq: maximum amplification gain}, one needs to repeatedly sweep over the entire $j\omega$-axis as $\alpha$ continuously increases; this is generally infeasible. 
In practice, one may perform a grid search over a restricted frequency range around the operating frequency; however, this scheme does not strictly guarantee stability.
In the following, we look for a lower bound on $\alpha_{\max}$.
	
\subsection{Lower Bound By Gershgorin Disc Theorem}
	
By Gershgorin disc theorem \cite[Th. 6.1.1]{horn2012matrix}, the eigenvalues of $\alpha\hat{\bH}(j\omega)$ are located in the union of discs: 
\begin{equation}
	\cG(\omega) \defeq \bigcup_{n\in[N]}\bigg\{z\in\C:|z| \leq \alpha\sum_{n'\neq n}|\hat{h}_{nn'}(j\omega)| \bigg\}.
\end{equation}
(Recall that we assume zero self-interference, i.e., $\hat{h}_{nn}=0,\forall n$.)
Therefore, a sufficient condition for the non-singularity of $\bI - \alpha\hat{\bH}(j\omega)$ is that $1\notin \cG(\omega)$, i.e.,
\begin{equation}
\label{eq: Gershgorin condition}
	\sup_{\omega} \max_{n\in[N]} \bigg\{\alpha \sum_{n'\neq n}|\hat{h}_{nn'}(j\omega)|\bigg\} < 1.
\end{equation}
This gives the following lower bound:
\begin{equation}
\label{eq: Gershgorin lower bound}
	\alpha_{\max} \geq \alpha_{\textup{G}} \defeq \inf_{\omega} \min_{n\in[N]} \frac{1}{\sum_{n'\neq n}|\hat{h}_{nn'}(j\omega)|}.
\end{equation}
Notice that $\{|\hat{h}_{nn'}(j\omega)|\}$ corresponds to the inter-repeater channel \emph{amplititude} gains. 
We have the following observation.
\begin{remark}
	It is the sum of the channel [amplitude] gains that matters, not the sum of the channel [power] gains. 
	In the worst case, the positive feedback combines constructively, in-phase, as if the repeaters formed a coherent antenna array.
\end{remark}

In what follows, we make case studies for some particular deployments of repeaters.
We consider isotropic antennas and  \gls{los} channels,
\begin{equation}
	\hat{h}_{nn'} = \sqrt{\beta_{nn'}} e^{-s \tau_{nn'}},
\end{equation}
where the channel power gain $\beta_{nn'}$ and the propagation delay $\tau_{nn'}$ are determined by the propagation distance $d_{nn'}$.
Notice that \gls{los} channels represent the worst case in our stability analysis, in the sense that inter-repeater interference becomes the strongest under \gls{los} propagation. 

\section{Case 1: Two Repeaters}
\label{sec: two-repeater}

Consider the case of  $N=2$ interacting repeaters. 
Due to channel reciprocity, we have
$
	\hat{h}_{12} = \hat{h}_{21} = \sqrt{\beta} e^{-s\tau}.
$
The inter-repeater channel transfer function matrix is 
\begin{equation}
	\hat{\bH}_: = 
	\begin{bmatrix}
		0 & \sqrt{\beta} e^{-s\tau} \\
		\sqrt{\beta} e^{-s\tau} & 0
	\end{bmatrix};
\end{equation}
\begin{flalign}
	& \mbox{therefore,} & \det(\bI - \alpha \hat{\bH}_:(j\omega)) = 1 - \alpha^2\beta e^{-j2\omega\tau}. & &
\end{flalign}

The plot of $\det(\bI - \alpha\hat{\bH}_:(j\omega))$ traces out a circle of radius $\alpha^2\beta$ centered at $s=1$ anticlockwise and periodically for every change  $\omega$ by of $\pi/\tau$.
Therefore, the two-repeater system is stable if and only if $\alpha^2\beta < 1$, indicating $\alpha_{\max} = 1/\sqrt{\beta}$.
Notice that the lower bound \eqref{eq: Gershgorin lower bound} is tight for this case.

For this particular case, we can also easily analyze the system in the time domain. 
First use \eqref{eq: signal in the time domain} to obtain
\begin{align}
	y_1(t) =& \alpha x_1(t) + \alpha\sqrt{\beta} y_2(t-\tau) \label{eq: time1}\\
	y_2(t) =& \alpha x_2(t) + \alpha\sqrt{\beta} y_1(t-\tau) \label{eq: time2}.
\end{align}
Recursively using   \eqref{eq: time1} and \eqref{eq: time2} yields
\begin{align}
	y_1(t) =& \alpha x_1(t) * p(t) + \alpha^2 \sqrt{\beta} x_2(t) * p(t-\tau)\\
	y_2(t) =& \alpha x_2(t) * p(t) + \alpha^2 \sqrt{\beta} x_1(t) * p(t-\tau),
\end{align}
where $p(t)$ is a ``decaying impulse train''
\begin{equation}
	p(t) \defeq \sum_{k=0}^\infty (\alpha^2\beta)^k \delta(t-2k\tau)
\end{equation}
representing the ``ping-pong'' effect of the loopback interference. 
The system is stable as long as $\alpha^2\beta < 1$ so that these ping-pongs decay exponentially fast to zero.

\section{Case 2: $N$ Repeaters on A Circle}

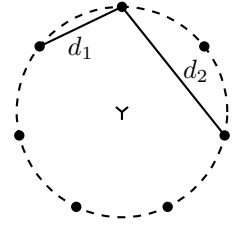
\begin{figure}
	\centering
	\begin{tikzpicture}[scale=0.7]
  	\draw[dashed, thick] (0,0) circle (2cm);
  
 	 \pgfmathsetmacro{\anglestep}{360/7}
  	\pgfmathsetmacro{\sinSexty}{sin(50)}
  	\pgfmathsetmacro{\sinThirty}{sin(40)}
  
 	 \foreach \i in {0,1,2,3,4,5,6}
    {
      \pgfmathsetmacro{\angle}{\i*\anglestep+90}
      \draw[fill] (\angle:2cm) circle (2.5pt) node[anchor=center] (dot\i) {};
      \ifnum\i=0
        \coordinate (firstDot) at (\angle:2cm);
      \fi
      \ifnum\i=1
        \coordinate (secondDot) at (\angle:2cm);
      \fi
      \ifnum\i=5
        \coordinate (fourthDot) at (\angle:2cm);
      \fi
    }
    \draw[thick] (firstDot) -- (secondDot) node[midway, below] {$d_1$};
    \draw[thick] (firstDot) -- (fourthDot) node[midway, right] {$d_2$};
    
  	\draw[thick] (-0.15*\sinSexty,0.15*\sinThirty) -- (0,0); 
  	\draw[thick] (0.15*\sinSexty,0.15*\sinThirty) -- (0,0); 
  	\draw[thick] (0,-0.15) -- (0,0); 
  
	\end{tikzpicture}
	\caption{Repeaters equally spaced on a circle.}
	\label{fig: circular network}
	\vspace{-0.5cm}
\end{figure}

Consider an odd number of repeaters, i.e.,  $N=2K+1$ for some positive integer $K$, uniformly spread out on a circle with radius $R$. 
(The analysis carries over to an even number of repeaters with slight changes. We omit that case for brevity.)
The source is located at the center of that circle. 

With a slight abuse of notation, we define 
\begin{equation}
	d_k \defeq 2R \sin(k\pi/N)
\end{equation}
and denote the channel power gain and the delay at distance $d_k$ as $\beta_k$ and $\tau_k$, respectively. 
By indexing the repeaters in the clockwise order, the channel between repeaters $n$ and $n'$ is
\begin{equation}
\label{eq: k-hop los channel}
	\hat{h}_k \defeq \sqrt{\beta_k} e^{-s \tau_k}
\end{equation}
\begin{flalign}
	& \mbox{with} & k = \min\{|n-n'|,N-|n-n'|\}. & &
\end{flalign}
See Fig. \ref{fig: circular network} for a graphical illustration.

\subsection{Eigenvalues of $\alpha\hat{\bH}$}

The inter-repeater channel transfer function matrix $\hat{\bH}_\circ$ is (complex-valued) symmetric circulant
	\begin{equation}
		\hat{\bH}_\circ = 
		\begin{bmatrix}[0]
			0 & \hat{h}_1 & \cdots & \hat{h}_N & \hat{h}_N & \cdots & \hat{h}_1 \\
			\hat{h}_1 & 0 & \ddots & & \ddots & \ddots &   \\
			\vdots & \ddots & \ddots \\
			\hat{h}_N \\
			\hat{h}_N & \ddots \\
			\vdots & \ddots\\
			\hat{h}_1
		\end{bmatrix}.
	\end{equation}
	Therefore, the loop transfer function matrix $\alpha\hat{\bH}_\circ$ is symmetric circulant with each column being a circulant permutation of 
	\begin{equation}
		\bv \defeq [0, \alpha\hat{h}_1,\cdots, \alpha\hat{h}_N, \alpha\hat{h}_N, \cdots, \alpha\hat{h}_1]^\transp.
	\end{equation}
	We can therefore denote $\alpha\hat{\bH}_\circ$ by $\Circ(\bv)$.
	
	It is well known that the eigenvalues of the circulant matrix $\Circ(\bv)$ are obtained by the \gls{dft} of $\bv$, i.e., the $n$-th eigenvalue is 
	\begin{equation}
	\label{eq: eigenvalues circular network}
	\begin{aligned}
		\lambda_n(\alpha \hat{\bH}_{\circ}) =& \sum_{i=0}^{N-1}[\bv]_n e^{-j2\pi i (n-1)/N}  \\
		=& 2\alpha\sum_{k\in[K]} \hat{h}_k\cos \frac{2\pi k (n-1)}{N}.
	\end{aligned}
	\end{equation}
	By substituting \eqref{eq: k-hop los channel} into \eqref{eq: eigenvalues circular network}, we obtain the $n$-th eigenvalue at frequency $\omega$ as 
	\begin{equation}
	\label{eq: eigenvalues circular network 2}
		\lambda_n(\alpha,\omega) = 2\alpha \sum_{k\in[K]} \left(\sqrt{\beta_k}\cos \frac{2\pi k (n-1)}{N}\right) e^{-j\omega\tau_k}
	\end{equation}
	The maximum amplification gain in this case becomes
	\begin{equation}
		\alpha_{\max} = \min \left\{\alpha > 0: \inf_{\omega} \min_{n\in[N]} |\lambda_n(\alpha,\omega) - 1| = 0  \right\}.
	\end{equation}
	Notice that even in this special case where the eigenvalues of $\alpha\hat{\bH}$ are obtainable in 
	closed form as a weighted sum of the complex exponentials $\{e^{-\omega\tau_k}\}$, determining whether $\lambda_n(\alpha,\omega) - 1$ has a root for an arbitrary $\alpha$ is still computationally difficult.
	It appears that only in the two-repeater case in Section \ref{sec: two-repeater}, it is tractable to find $\alpha_{\max}$ analytically.
	
\subsection{Numerical Stability Analysis}
\label{subsec: freq-sweeping}

We check the stability of the circular repeater network in Fig. \ref{fig: circular network} numerically.
The cell radius is taken to be $R=1000$ meters.
We consider the free-space propagation model where the channel power gain between two repeaters at distance $d$ is
\begin{equation}
	\beta(d) = \frac{\lambda^2}{(4\pi)^2} \frac{1}{d^2},
	\label{eq: channel-gain}
\end{equation}
with $\lambda$ being the wavelength at the carrier frequency, which is set to 2 GHz.
The propagation delay is $\tau(d) = d/c$, where $c\approx 3\times10^8$ m/s is the speed of light. 
For each value of $\alpha$, we calculate 
\begin{equation}
	\min_{\omega\in\Omega} \min_{n\in[N]} |\lambda_n(\alpha,\omega) - 1|
\end{equation}
as a measure of stability, where $\Omega$ is a discrete set of angular frequencies corresponding to a 20 MHz frequency band centered around the carrier frequency with 100 Hz spacing.

\begin{figure}[t]
	\centering
	\includegraphics[width=7cm]{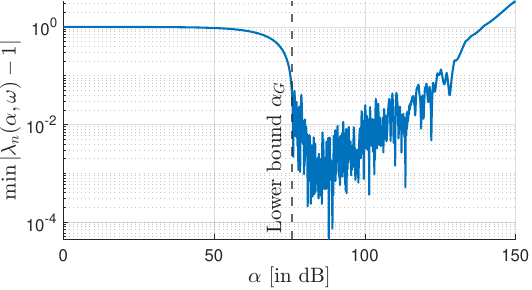}
	\caption{Stability test for repeaters on a circle.}
	\label{fig: numerical-circular}
	\vspace{-0.5cm}
\end{figure}

The result with $N=15$ repeaters is shown in Fig. \ref{fig: numerical-circular}.
One can observe that the lower bound in \eqref{eq: Gershgorin lower bound} obtained by the Gershgorin disc theorem quite accurately captures the transition point at which the system starts to become unstable -- hence, the bound is quite tight in this case.

\subsection{Coverage Extension}

\begin{figure}[t]
	\centering
	\includegraphics[clip, trim=0cm 1.8cm 0cm 0cm, width=6cm]{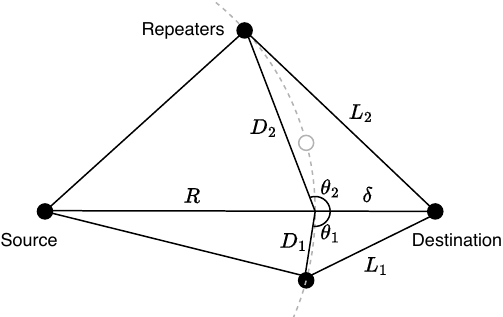}
	\caption{Illustration of coverage extension.}
	\label{fig: coverage extension}
\end{figure}

To get a first-order insight into the potential for using repeaters for coverage extension, in terms of ``pushing the
cell boundary,'' we consider a destination located $R+\delta$ meters away from the source; see 
 Fig. \ref{fig: coverage extension}. 
The source is assumed to be able to provide wireless coverage up to the distance $R$, without repeaters turned on.
We want to check how large amplification gain $\alpha$ is needed for the repeaters to extend the coverage to the destination at the distance $R+\delta$.
We assume that the destination has equal distances to the two closest repeaters. 

When there are $N=2K$ repeaters (again, the analysis can be easily extended to an odd number of repeaters), the distances from the repeaters to the destination are $L_1,\cdots,L_K$ (each has multiplicity 2), where
\begin{equation}
	\label{eq: distance from repeater to destination}
	L_k= \sqrt{\delta^2 + D_k^2 - 2\delta D_k \cos\theta_k}
\end{equation}
with
\begin{equation}
	\label{eq: D}
	D_k = 2R\sin \frac{(2k-1)\pi}{2N} ~ \text{and} ~ \theta_k = \frac{(2k-1)\pi}{2N} + \frac{\pi}{2}.
\end{equation}

The minimum $\alpha$ to achieve coverage extension $\delta$ satisfies
\begin{equation}
	\beta(R+\delta) + 2\alpha^2 \beta(R) \sum_{k\in[K]}\beta(L_k)  = \beta(R),
\end{equation}
\vspace{-0.2cm}
\begin{flalign}
	\label{eq: minimum alpha for coverage}
	\mbox{or equivalently,} && \alpha = \sqrt{\frac{\beta(R) - \beta(\delta+R)}{2\beta(R)\sum_{k\in[K]}\beta(L_k)}}, &&
\end{flalign}
where we used the free-space channel gain in \eqref{eq: channel-gain}.

\begin{figure}
	\centering
	\includegraphics[width=7cm]{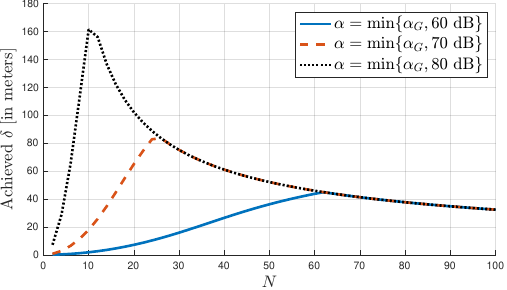}
	\caption{Achieved coverage extension with different $N$.}
	\label{fig: delta}
	\vspace{-0.5cm}
\end{figure}

By using \eqref{eq: minimum alpha for coverage}, we can (approximately) test whether a coverage extension $\delta$ is possible by comparing the required $\alpha$ with $\alpha_G$ in \eqref{eq: Gershgorin lower bound} and the power constraint.
In Fig. \ref{fig: delta} we show the coverage extension achieved by $N$ repeaters when they use the amplification gain $\alpha = \min\{\alpha_G,\gamma\}$, where $\gamma \in \{60,70,80\}$ dB represents the transmit power constraint of the repeaters. 
The other parameters are set to be the same as in Section \ref{subsec: freq-sweeping}.
We can observe that as the number of repeaters increases, the achieved coverage extension is first restricted by the power constraint and then by the stability requirement. 
The optimal deployment of repeaters depends on the power constraint: 
when the repeaters transmit with high power, they must be sparsely deployed to avoid instability.
Quantitatively, from  Fig. \ref{fig: delta}, the coverage extension that can be obtained is quite small.

\section{Case 3: Repeaters on A Grid}

\begin{figure}
	\centering
	\includegraphics[width=7cm]{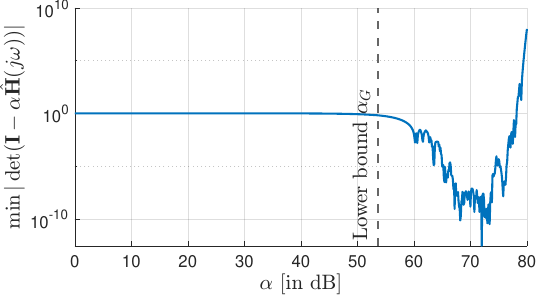}
	\caption{Stability test for repeaters on a grid.}
	\label{fig: freq-sweeping-grid}
	\vspace{-0.2cm}
\end{figure}

We next consider a case when the repeaters are uniformly located on a grid in a square-shaped cell of size $2\times 2$ kilometers. 
  Fig. \ref{fig: freq-sweeping-grid} shows  the value of 
\begin{equation}
	\min_{\omega\in\Omega}|\det(\bI - \alpha\hat{\bH}(j\omega))|,
\end{equation}
as a measure of stability when the distance between two adjacent repeaters is 200 meters.
The channel model and the other parameters are the same as in Section \ref{subsec: freq-sweeping}.
We can observe that the lower bound obtained from the Gershgorin disc theorem again predicts the transition point rather accurately.

\begin{figure}
	\centering
	\includegraphics[width=7cm]{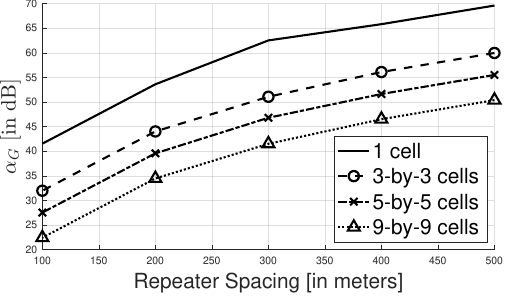}
	\caption{Values of $\alpha_G$ with different repeater spacings.}
	\label{fig: num-cells}
	\vspace{-0.5cm}
\end{figure}

To check how densely the repeaters can be deployed and how many cells can be in operation simultaneously, we plot the value of the lower bound $\alpha_G$ with different repeater spacings and different numbers of cells in Fig. \ref{fig: num-cells}. 
We observe that the system stability is not only constrained by how densely the repeaters are deployed in each cell but also by the number of cells that are simultaneously operated. 
To prevent destructive positive feedback, one may have to coordinate the repeater operation in different cells.

\section{Conclusions}

We investigate a wireless network where multiple repeaters are operating simultaneously and interacting with each other. 
The interaction forms a positive feedback loop which could, depending on the amplification gains, become unstable. 
To analyze this problem, we used the Gershgorin disc theorem to obtain a lower bound on the maximum amplification gain that the repeaters can safely use.
This analysis reveals that the maximum amplification gain is constrained by the sum of the inter-repeater channel amplitude gains, rather than the sum of the inter-repeater path losses.
Through case studies, we show that: 1) the coverage extension that can be achieved by the repeaters can be severely restricted by the stability requirement; 2) how densely the repeaters should be deployed depends on both the power and the stability constraints; 3) in multi-cell systems, repeaters operating in different cells need to be properly coordinated to avoid stability problems; and 4) in large-scale systems, the stability criterion can be easily wrecked by a locally poor deployment -- it is enough that only two of the repeaters are too closely located. Importantly, the stability criteria depend only on the repeater deployment and channels, and have no dependence on the locations, or transmit powers, of the user terminals or the base station.

\emph{Future Directions:}
The  analysis may be extended to consider the effects of noise injected by the repeaters. 
Also, the repeaters may have different amplification gains such that $\alpha$ becomes dependent on $n$, and introduce  phase shifts such that $\alpha$ becomes complex-valued (or even small time-delays such that $\alpha$ becomes  dependent on $\omega$), which was not considered here.
Moreover, our initial case studies assumed isotropic antennas. The effects of different antenna patterns remain to be analyzed.

\bibliographystyle{ieeetr}
\bibliography{ref}

\end{document}